\documentclass{article}
\usepackage{RR}
\RRNo{6238}

\usepackage[english]{babel}
\usepackage{cite}
\usepackage{graphics}
\usepackage[usenames]{color}
\usepackage{subfigure}
\usepackage[ruled,linesnumbered,algonl]{algorithm2e}
\usepackage{float,afterpage}
\usepackage{comment}
\usepackage{amssymb}
\usepackage{subfigure}
\usepackage{simplemargins}
\usepackage{ifthen}
\usepackage{url}
\usepackage{paralist}
\usepackage{caption}
\newcommand{\reffig}[1]{Fig.~\ref{#1}}
\newcommand{\refsubfig}[1]{\ref{#1}}

\newcommand{\refsec}[1]{section~\ref{#1}}

\newcommand{\LONG}[1]{
{\em \color{red} #1}
}

\newcommand{\cu}[1]{\mathcal{#1}}

\newcommand{\XXX}[1]{}
\newcommand{\remove}[1]{}

\newcommand{\ForgetXXX}[1]{}

\graphicspath{{./}{./Graph/}}

\newcommand{\DS}{\Delta{}S}

\newcommand{\Cmin}{C_\mathrm{min}}

\newcommand{\Ecost}{E_\mathrm{cost}}

\renewcommand{\thesubfigure}{Fig.~\thefigure.\arabic{subfigure}}
\makeatletter
  \renewcommand{\@thesubfigure}{\thesubfigure:\space}
  \renewcommand{\p@subfigure}{}
\makeatother

\usepackage{xspace}
\usepackage{ifthen}

\usepackage{epsfig}

\newcommand{\thetitle}[0]{ Heuristics for Network Coding in Wireless Networks }

\newcommand{\authorlist}[0]{Song Yean Cho, C\'edric Adjih, Philippe Jacquet}

\begin{document}


\RRetitle{
Heuristics
for Network Coding in Wireless Networks} 
\RRtitle{
Heuristiques
pour le codage de r\'eseau dans les r\'eseaux sans fil}

\RRdate{Juin 2007}

\RRauthor{\authorlist}
\authorhead{\authorlist}
\titlehead{\thetitle}

\RRtheme{\THCom}
\RRprojet{HIPERCOM}
\URRocq

\RRresume{
Le multicast est un enjeu central pour les nouvelles architectures
de r\'eseaux sans fil, tels que les r\'eseaux maill\'es ``mesh'',
\`a cause de son co\^ut substanciel en termes de bande passante. \\
Dans ce rapport, nous \'etudions une m\'ethode de multicast
sp\'ecificique: la diffusion \`a tout le r\'eseau, dans un r\'eseau
sans fil multi-sauts. Plus pr\'ecisement, la diffusion que nous
consid\'erons est fond\'ee sur l'utilisation du codage de r\'eseau,
une m\'ethode prometteuse pour r\'eduire le co\^ut. Nos pr\'ec\'edents
travaux ont montr\'e que la performance du codage de r\'eseaux avec une
heuristique simple \'etait assymptotiquement optimale : chaque
transmission est utile \`a presque tous les destinataires.
C'est vrai pour de grands r\'eseaux homog\`enes du plan Euclidien.

Mais pour des r\'eseaux plus petits, ou inhomog\`enes, des heuristiques
suppl\'ementaires sont requises. Ce rapport propose de telles heuristiques
(de choix de d\'ebit) pour la diffusion par codage de r\'eseau. Nos
heuristiques ont pour objectif d'utiliser seulement des informations
simples concernant la topologie locale. Nous d\'etaillons la logique
des heuristiques, et par des r\'esultats exp\'erimentaux,
nous illustrons le comportement de ces heuristiques, et mettons
en \'evidence leur excellente performance.
}

\RRabstract{ 

Multicast is a central challenge
for emerging multi-hop wireless architectures such as wireless mesh
networks, because of its substantial cost in terms of bandwidth.\\
In this report, we study one specific case of multicast:
broadcasting, sending data from one source to all nodes, in a
multi-hop wireless network. The broadcast we focus on is based on 
network coding,
a promising avenue for reducing cost; previous work of ours
showed 
that the performance of network coding with simple
heuristics is asymptotically optimal: each
transmission is beneficial to nearly every receiver. 
This is for homogenous and large networks of the plan.

But for small, sparse or for 
inhomogeneous networks, some additional
heuristics are required. This report proposes such additional new
heuristics (for selecting rates) 
for broadcasting with network coding. Our heuristics
are intended to 
use only simple local topology information. We detail the
logic of the heuristics, and with experimental results, 
we illustrate the behavior of the heuristics, and demonstrate
their excellent performance.
}

\RRmotcle{
sans fil, diffusion, codage de r\'eseau, hypergraph, coupe minimale,
heuristique, s\'election de d\'ebit
}

\RRkeyword{
wireless, broadcast, network coding, hypergraph, min-cut,
heuristics, rate selection, subgraph selection
}

\makeRR


\section{Introduction}
The confirmed success of wireless networks has made wireless
communication ubiquitous. One of the predicted use of wireless
networks is multicast, which could be used for multimedia content
diffusion, video conference, software distribution, and a number
of other applications. However, in multi-hop networks such as
wireless mesh networks, multicast transmissions incur a
substantial cost by simple virtue of requiring relaying over
several forwarders, in order to cover every destination.
For this reason efficient techniques for multicasting are of prime
interest. One of them is the recently proposed method, network
coding. Network coding was introduced by the seminal work of
\cite{Bib:ACLY00} as a new paradigm 
where 
intermediate nodes mix
information from different flows (different bits or different
packets). We use network coding 
specifically for broadcasting rather then general multicasting in wireless
multi-hop networks.

The problem that we are addressing is \emph{efficient broadcast}:
\begin{itemize}
\item Broadcast packets from one source to all nodes, with the
minimum number of transmissions.
\end{itemize}

Without network coding, finding the optimal broadcasting is an NP
complete problem~\cite{Bib:CHE02}, but a number of
heuristics exist for efficient broadcasting
such as MPR-flooding \cite{Bib:olsr}
or techniques
based on connected dominating sets 
\cite{Bib:GK96,Bib:AJV05}. 
But with network coding, the optimal broadcasting can be found in
polynomial time. Finding an optimal
solution (subgraph selection) consists in 
finding the coding nodes and their optimal rates
\cite{Bib:LMKE05,Bib:HKMKE03,Bib:LMKE07}. This
problem can be formulated as a linear program, which can be solved
in polynomial time \cite{Bib:WCK05,Bib:LRMKKHAZ06}, and possibly
in a distributed fashion \cite{Bib:LRMKKHAZ06}.

However, we adopt a different, even simpler, approach:
previous
work \cite{Bib:ACJ07,Bib:RR-ACJ07} has shown that a simple
heuristic for selecting rates
could achieve asymptotically the optimal efficiency for
homogeneous large and dense wireless networks of the plane --- and
also that, noticeably, it would outperform methods without network
coding. This is asymptotically true for homogeneous networks, but
the heuristic needs adjustments for less homogeneous, smaller or
sparser networks. The adjustment is the topic of this report.
Our key contributions are the following:
\begin{itemize}
\item We propose an improved heuristic for rate selection,
inspired by~\cite{Bib:FWB06}. It requires only local topology
information: knowledge of two-hop neighbors.
\item We empirically study its performance on representative graphs with
different densities and different sizes. We investigate and
explain the variation of the performance, and also compare it to
other techniques (including without using network coding).
\end{itemize}

The rest of the paper is organized as follows: \refsec{sec:definition}
details the network model and definitions, \refsec{sec:heuristics}
describes the
heuristics, \refsec{sec:simulations} analyzes performance
with experimental results and \refsec{sec:conclusion} concludes. 

\section{Definitions}
\label{sec:definition}

\subsection{Network Model}

In this report, we study the problem of broadcasting from one
source to all nodes. 
We will focus on getting a preliminary
idea of the performance of our heuristics for wireless networks.
Hence, we will assume an ideal wireless model in this report
(and realistic models would be subject of future work):
lossless wireless transmissions without collisions or
interferences. We also assume that each node of the network is
operating below its maximum transmission capacity.

\begin{figure}[htp]
\subfigure[random unit disk graph]{\label{fig:random-es}
\includegraphics[width=5cm]{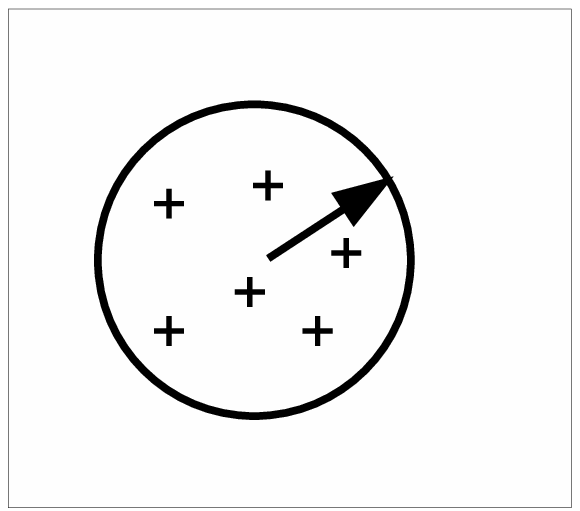}
}\hspace{.1in}
\subfigure[lattice graph]{ \label{fig:lattice-es}
\includegraphics[width=.25\textwidth]{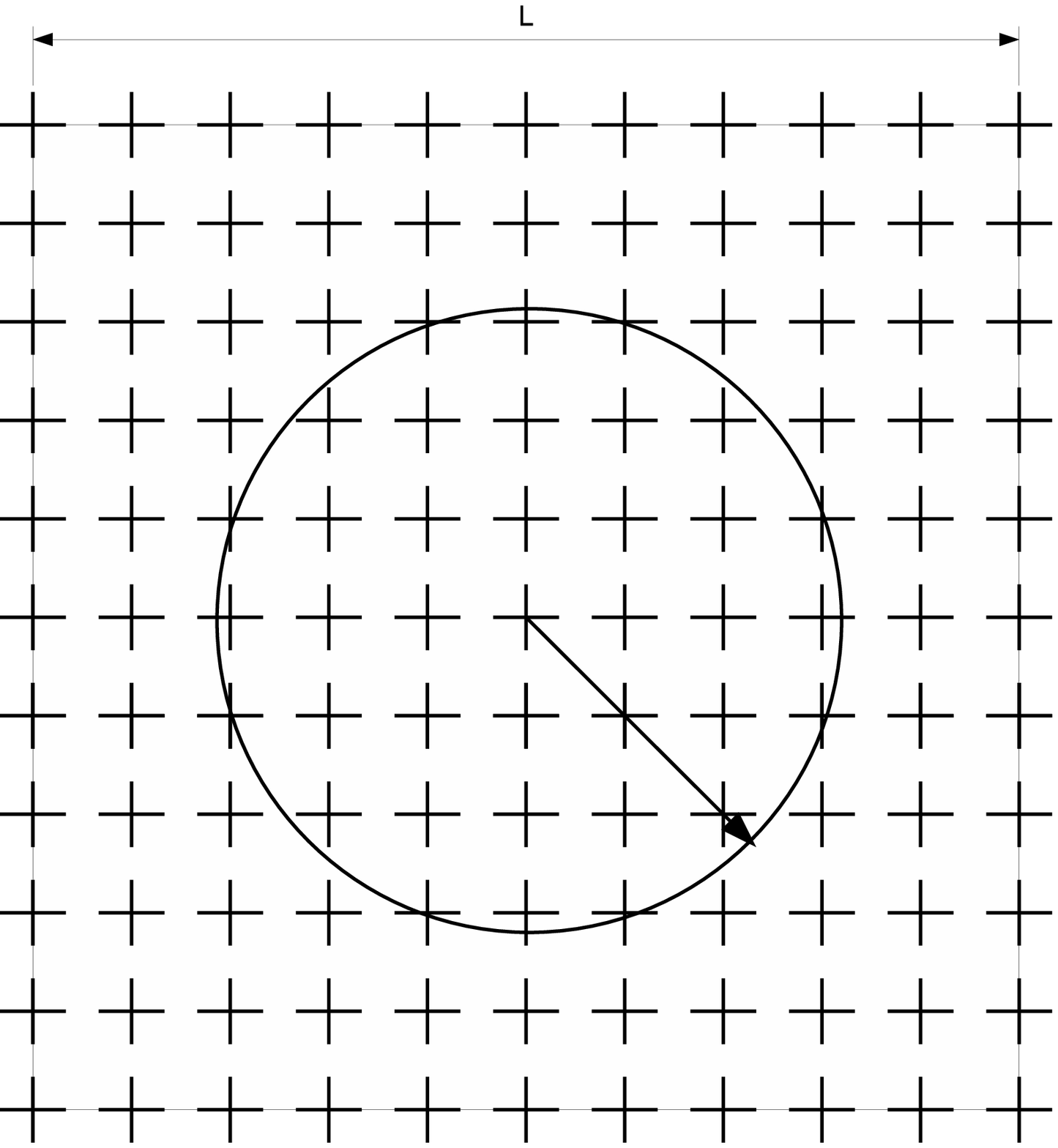}
}\hspace{.1in}
\subfigure[on a torus]{\label{fig:random-torus}
\includegraphics[width=4cm]{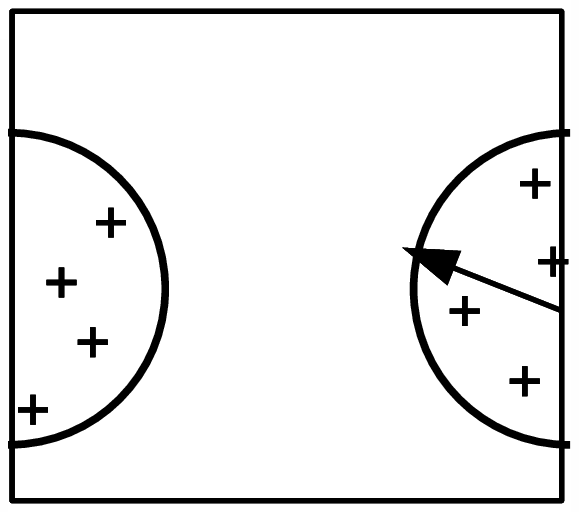}
}%
\vspace{-4mm}
\caption{Network Models}
\label{fig:unit-disk}
\end{figure}

In an idealized model, multi-hop wireless networks can be
modeled as \emph{unit disk graphs}
of the plane, where two nodes are neighbors if their distance is
lower than a fixed radio range as seen in~\reffig{fig:unit-disk}.
In addition, in wireless
networks, the \emph{wireless broadcast advantage} is used: each
transmission is overheard by several nodes. As a result the graph
is in reality a \emph{(unit disk)
hypergraph}. 
Precisely, we consider the following networks: unit disk
[hyper]graphs where nodes are either distributed randomly
(\refsubfig{fig:random-es}) or more regularly organized in a lattice
(\refsubfig{fig:lattice-es}). In addition, in both cases, we also
consider their variants where the network is a torus,
with wrap-around connections in both
the $x$ and $y$ directions as on~\refsubfig{fig:random-torus}.
%

\subsection{Notation}

We will consider the following items in each instance of the
several types of graphes in section~\ref{sec:definition},
and use the following notation in the rest of the report:
\begin{itemize}
\item {\bf Nodes:} $\cu{V} $, set of
vertices (nodes) of the graph
\item {\bf Number of nodes:} $N$
\item {\bf Expected/average number of neighbors:} $M$
\item {\bf Hyperedge {}:}  $H_v$, $H_v \subset \cu{V}$ 
is the subset of nodes which are reached by one transmission
of node $v$ (the neighbors of $v$)
\item {\bf Rate:} Each node $v$ retransmits coded packets with a fixed rate $C_v$.
\item {\bf Cost per broadcast:} $E_\mathrm{cost}$; defined is
\refsec{sec:performance}

\item {\bf Optimal cost per broadcast :} $E_\mathrm{optimal}$
(\refsec{sec:performance})

\item {\bf Min-cut of the source $s$, for broadcast to the entire network:}
$C_\mathrm{min}(s)$; see \refsec{ch:min-cut}

\end{itemize}

\subsection{Performance}
\label{sec:performance}

Because we focus on broadcasting, our approach uses a simplified
subgraph selection, where every node is a coding node. The
solution for the simplified subgraph consists of defining the rate
of each node -- by the heuristic. 

To measure the performance of the heuristics, we use the number of
transmissions per broadcast as the metric for the cost. We consider:

\begin{itemize}
\item the number of retransmissions from every node per unit time
(directly given by selected rate).
\item the number of packets
successfully broadcasted from the source to the entire network per
unit time;
it is the achievable broadcast rate. 
\end{itemize}
By dividing the number of retransmissions by the number of packets
successfully broadcasted,
we obtain our metric for the cost per
broadcast and denote it $E_\mathrm{cost}$. For
reference, we will also use the cost of the optimal solution,
$E_\mathrm{optimal}$, obtained by solving the linear program as
presented in \cite{Bib:LRMKKHAZ06}.

The number of packets successfully broadcasted per the unit time ,
i.e. the maximum achievable broadcast rate is computed as the
\emph{min-cut} from the source to every destination in
the network considered as a hypergraph 
as defined in section~\ref{ch:min-cut}.

\subsection{Achievable Broadcast Rate: Min-cut}
\label{ch:min-cut}

A central result of the performance of network coding
in wireless networks  
gives the maximum broadcast (more generally: multicast) rate for a
source. It is given by the \emph{min-cut} from the source to each
individual destination of the networks, viewed as a hypergraph
\cite{Bib:min-cut-hypergraph}.

Let us consider the source $s$, and one of the
broadcast destinations $t \in \cu{V}$. The definition of a
\emph{$s$-$t$ cut} is: a partition of the set of vertices $V$ in
two sets $S$, $T$ such as $s \in S$ and $t \in T$. Let $Q(s,t)$ be
the set of such \emph{$s$-$t$ cuts}: $(S,T) \in Q(s,t)$.

We denote $\DS$, the set of nodes of $S$ which are neighbors of at
least one node of $T$ ; the \emph{capacity of the cut} $C(S)$ is
defined as the maximum rate between the nodes in $S$ and the nodes
in $T$:
\vspace{-3mm}
\begin{equation}
\label{eq:cut-capacity}  C(S) \triangleq \sum_{v \in \Delta{}S}
C_v
\end{equation}

The \emph{min-cut} between $s$ and $t$ is the cut of $Q(s,t)$ with
the minimum capacity. Let us denote $C_\mathrm{min}(s,t)$ its
capacity. 
From \cite{Bib:GDPHE04,Bib:DGPHE06,Bib:min-cut-hypergraph}, 
the maximum broadcast capacity is given by the minimum of capacity of
the min-cut of every destination $C_\mathrm{min}(s)$, with:
\vspace{-2mm}
$$C_\mathrm{min}(s,t) \triangleq \min_{(S,T) \in
Q(s,t)} C(S) ~~~\mathrm{and}~~~ C_\mathrm{min}(s) \triangleq
\min_{t \in \cu{V} \setminus \{s\}} C_\mathrm{min}(s,t)$$

Once computed, the performance given by the global min-cut
$\Cmin(s)$ can be achieved by different coding methods
(precisely or asymptotically).

\subsection{Previous Heuristic}
\label{sec:iron-only}

In \cite{Bib:ACJ07,Bib:RR-ACJ07} we proposed a simple heuristic,
where most nodes have the same rate except the source and some
nodes near the edge of the network (\emph{exceptional nodes}):
IREN/IRON (\emph{Increased Rate for Exceptional Nodes, Identical
Rate for Other Nodes}).

The simple heuristic achieved asymptotically near-optimal
efficiency for unit disk graphs in large dense homogenous networks
. In other words, the simple heuristic enables for every transmission
to bring \emph{innovative} information to almost every
receiver.

Let us reproduce the logic of the heuristic:
\begin{enumerate}
\item Assume that the every node has an identical retransmission
rate. Assume it is $1$, arbitrarily, e.g. one packet per second.
\item \label{step:receive} Then every node with $M$ neighbors can
receive $M$ coded packets per second. Assume that nearly all of
them innovative. \item Then the source should inject at least $M$
packets per a second. \item \label{step:border} An issue is the
nodes of near the border , because they have less neighbors - so
in order to be safe, their rate is set to $M$ as well (IREN).
\end{enumerate}

Within this framework, it was proven that the achievable broadcast
capacity, the min-cut, would be $M$ for lattice graphs, and
asymptotically $M$ for dense random unit-disk graphs. For large
lattice or random unit-disk graphs, the cost per broadcast would
also converge to the optimal.However the result is asymptotical; 
for a practical instance of a
graph, the cost of the ``increased rate'' of the border nodes could
be considerable, and unnecessary high.

An immediate way to alleviate the cost is to omit
step~\ref{step:border} of the reasoning entirely, ie, ignoring 
the issue of border, thus  using the following
heuristics:
\begin{itemize}
\item IRON only (Identical rate for other nodes): every node
retransmits with rate $1$, except for the source,
which transmits with rate $M$.
\end{itemize}

\begin{figure}[htp!]
\centering 
\subfigure[min-cut with IRON only, on a
lattice]{\label{fig:iron-lattice-es}
\includegraphics[width=5cm]{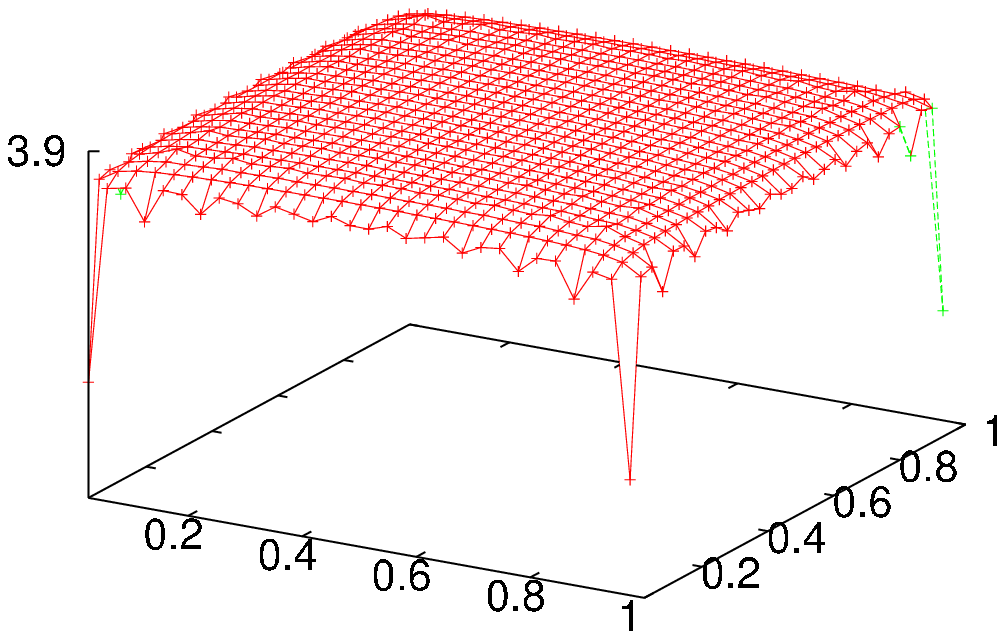}
}
\hspace{.1in}
\subfigure[min-cut with IRON only, on a random unit disk]
{\label{fig:iron-random-es}
\includegraphics[width=3cm,angle=270]{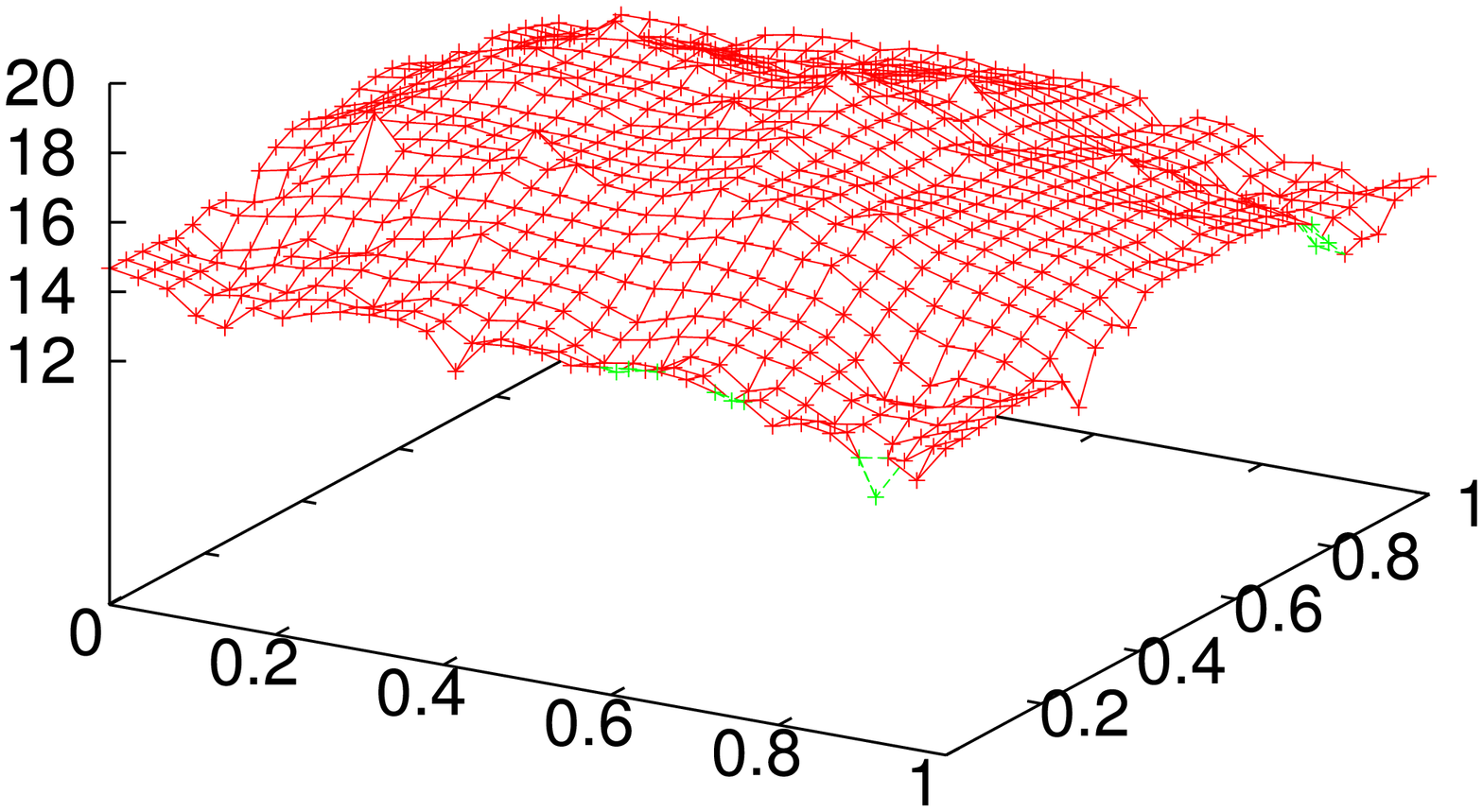}
}
\caption{Min-cut with IRON only}
\label{fig:iron-only}
\end{figure}

For illustrative purposes, we show the maximum achievable broadcast rate,
for two instances of the networks with the rates selected by
``IRON only'' on \reffig{fig:iron-only}.

The \reffig{fig:iron-only} represents the min-cut $C_\mathrm{min}(s,t)$ of
each node $t$ with respect to its position when the total number of nodes
is $N=400$. Both $x$ and $y$ axis
represent the position of each node on the plane;
the value on the vertical $z$ axis represents 
min-cut\footnote{for the randomly generated unit disk graph, there
is interpolation}. 
On a lattice, \refsubfig{fig:iron-lattice-es}, every node has $M=4$
neighbors, and the source is in the middle of the network. In the
random unit disk graph in \refsubfig{fig:iron-random-es} 
we select the node with the most number of
neighbors as the source, and generate graphs 
with density (the average number of nodes in a range)
equal to $M=20$. 

The min-cut was computed with the software library implementing
the maxflow computation algorithm from \cite{Bib:BK04} (and one
additional layer to model a directed hypergraph as a directed graph).
The optimizations for tree reuse from~\cite{Bib:KT05} were also used.

As seen in\reffig{fig:iron-only}, the nodes near the border 
have a min-cut much lower 
than nodes in the middle of the network ($4$ compared to $2$
for \refsubfig{fig:iron-lattice-es}).
From \refsec{ch:min-cut}, recall that maximum achievable broadcast
rate from the source, $C_\mathrm{min}(s)$, is the minimum of the 
min-cuts $C_\mathrm{min}(s,t)$ to each destination $t$.
Hence these nodes near the border become the bottleneck 
and dictate
a bound for the source rate much lower than what other nodes could 
have achieved.
Also in random unit-disk graph
\refsubfig{fig:iron-lattice-es}, notice the same phenomenom occurs
on the borders but also that the irregularity of number of neighbors 
results in irregularity of the min-cut (and hence potential lower
minimum min-cut).

These are reasons of inefficiency of IRON.

%
%


\subsection{The Proposed Heuristic: IR-MS}
\label{sec:heuristics}

As illustrated with the simple heuristics in
\refsec{sec:iron-only}, the lack of neighbors lets border nodes
and some other nodes receive less than $M$ packets per second and
these nodes cause a decrease of the performance. We name these
nodes \emph{starving nodes}

To alleviate the bottleneck from the starving nodes, their neighbors
compensate their starving by increasing the transmission rates with the
following heuristic, inspired by~\cite{Bib:FWB06}:
\begin{itemize}
\item IR-MS (Increased Rate for the Most Starving node): the rate
of
a node $v$ is set to $C_v$, with: \\
$C_v$ =  $\frac{M}{min_{u \in H_v}(|H_u|)}$, where $H_w$ is the
set of neighbors of $w$, and $M$ is the source rate.
\end{itemize}

This rate selection adjusts the step~\ref{step:receive} of the
reasoning presented in \refsec{sec:iron-only}, to the fact that
some nodes have less than $M$ neighbors. When IR-MS, if a node $v$ has less
than $|H_v| < M$ neighbors, it would still receive a packet rate
$\ge \frac{M}{|H_v|}$ from each of its $|H_v|$ neighbors, and hence
an overall rate at least equal to $M$. Notice that this does not
necessarily result in a min-cut
 $C_\mathrm{min}(s,v) \ge M$.

\begin{figure}[htpb]
\vspace{-5mm} \centering \subfigure[min-cut with IR-MS, on a
lattice]{ \label{fig:irms-lattice-es}
\includegraphics[width=6cm]{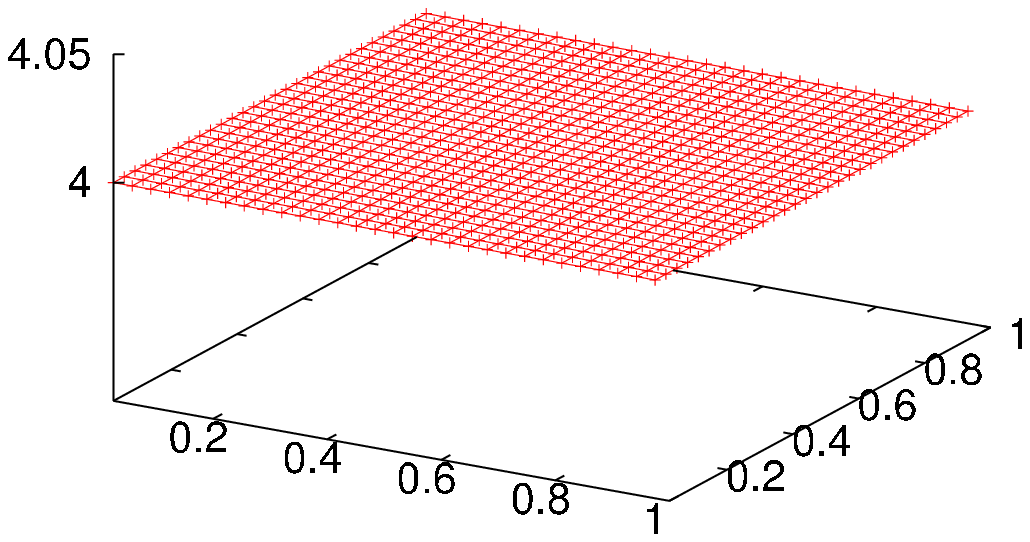}
}\hspace{.1in}
\subfigure[min-cut with IR-MS, on a random unit disk]{\label{fig:irms-random-es}
\includegraphics[width=5cm]{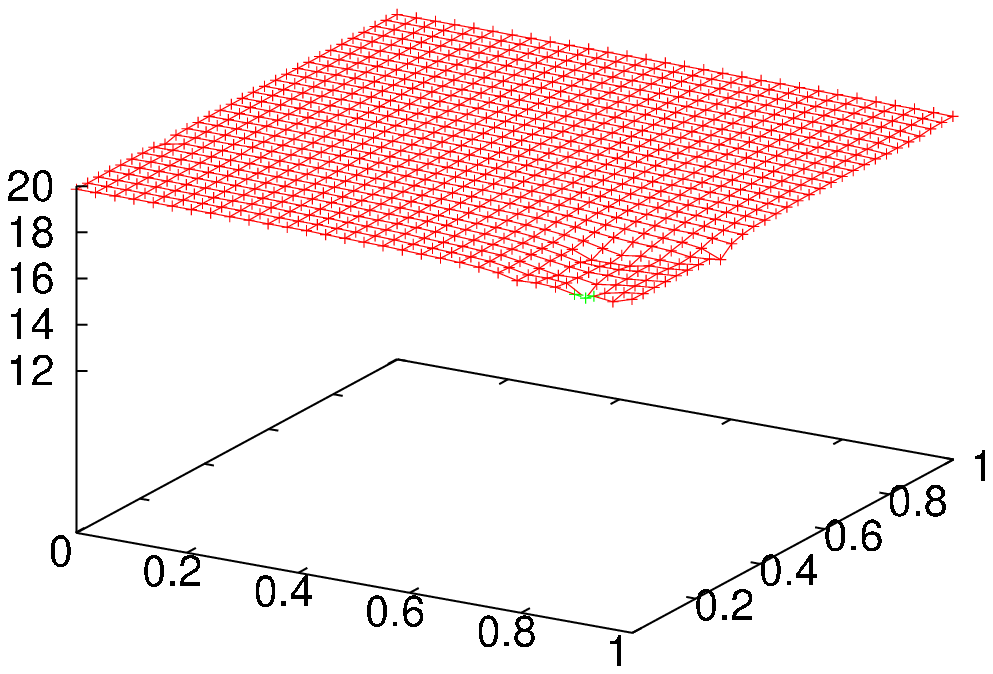}
}
\vspace{-5mm}
\caption{Min-cut with IR-MS}
\label{fig:irms}
\end{figure}
The \reffig{fig:irms} represents the value of the min-cut with
the rate selection of IR-MS, with the same topologies and same
parameters as for IRON only, in previous \reffig{fig:iron-only}. It 
appears that for these topologies,
the min-cut has the value which is  much closer to the targeted
value $M$. The increased rates of IR-MS at neighbors of
bottleneck may bring additional costs. However this additinal cost is 
relatively much lower than the increase of min-cut, thus the performance is
overally improved. 
We will present more systematic
experiments on the performance of the heuristic IR-MS in the next
section.

%
%

\section{Experimental Results}
\label{sec:simulations}

\subsection{Efficiency on Different Types of Networks}

We first evaluate the efficiency
of the heuristic IR-MS,
with the relative cost w.r.t. the optimal:
$E_\mathrm{rel-eff} \triangleq \frac{E_\mathrm{optimal}}{E_\mathrm{cost}}$

One reference point is the approximative upper bound
in~\cite{Bib:RR-ACJ07}, the achievable performance without network
coding, which translates into:
$E_\mathrm{bound-rel-eff}^\mathrm{(no-coding)} \approx 0.609\ldots$.

\begin{figure*}[htpb]
\label{fig:random-unit-disk}
\centering
\subfigure[micut: cumulative distribution]{\label{fig:mincut-cmu-dis}
\includegraphics[width=.4\textwidth]{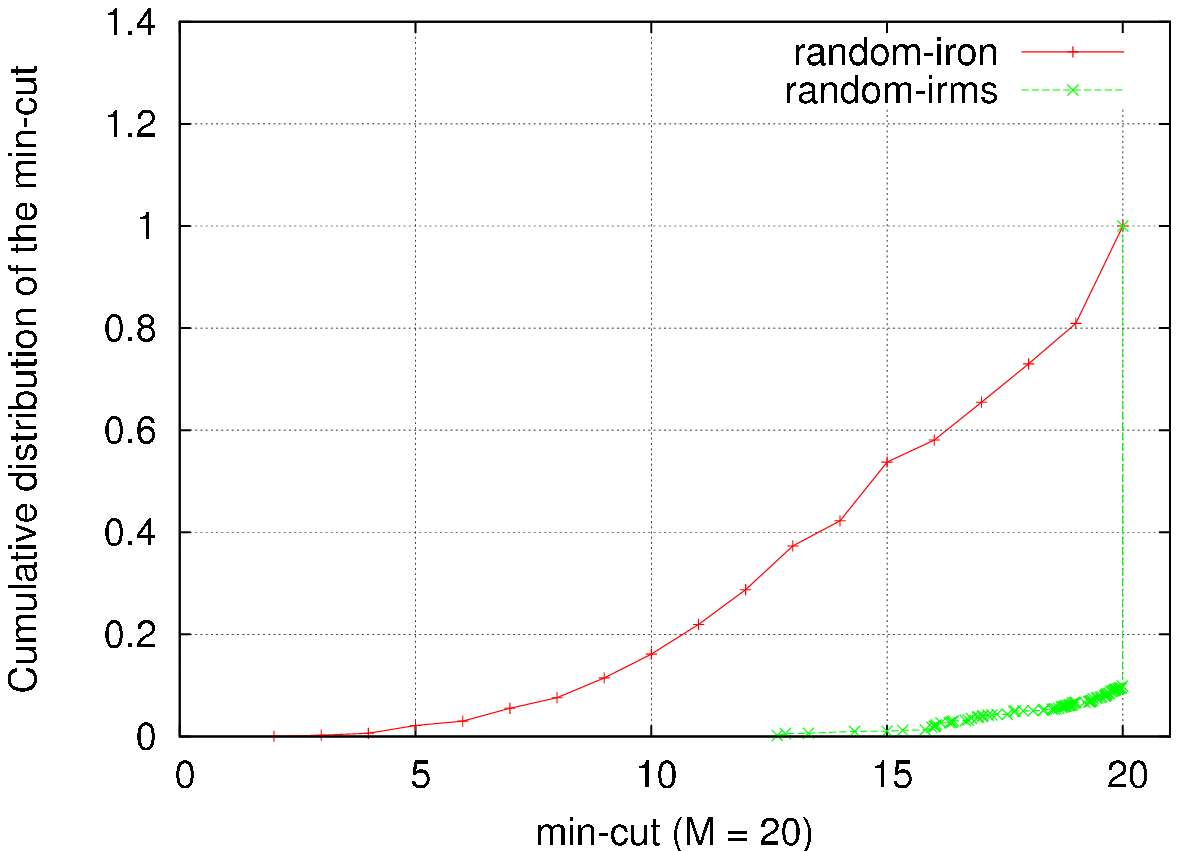}
}
\hspace{.1in}
\subfigure[mincut vs distance from the border]{\label{fig:mincut-mu-dis}
\includegraphics[width=.4\textwidth]{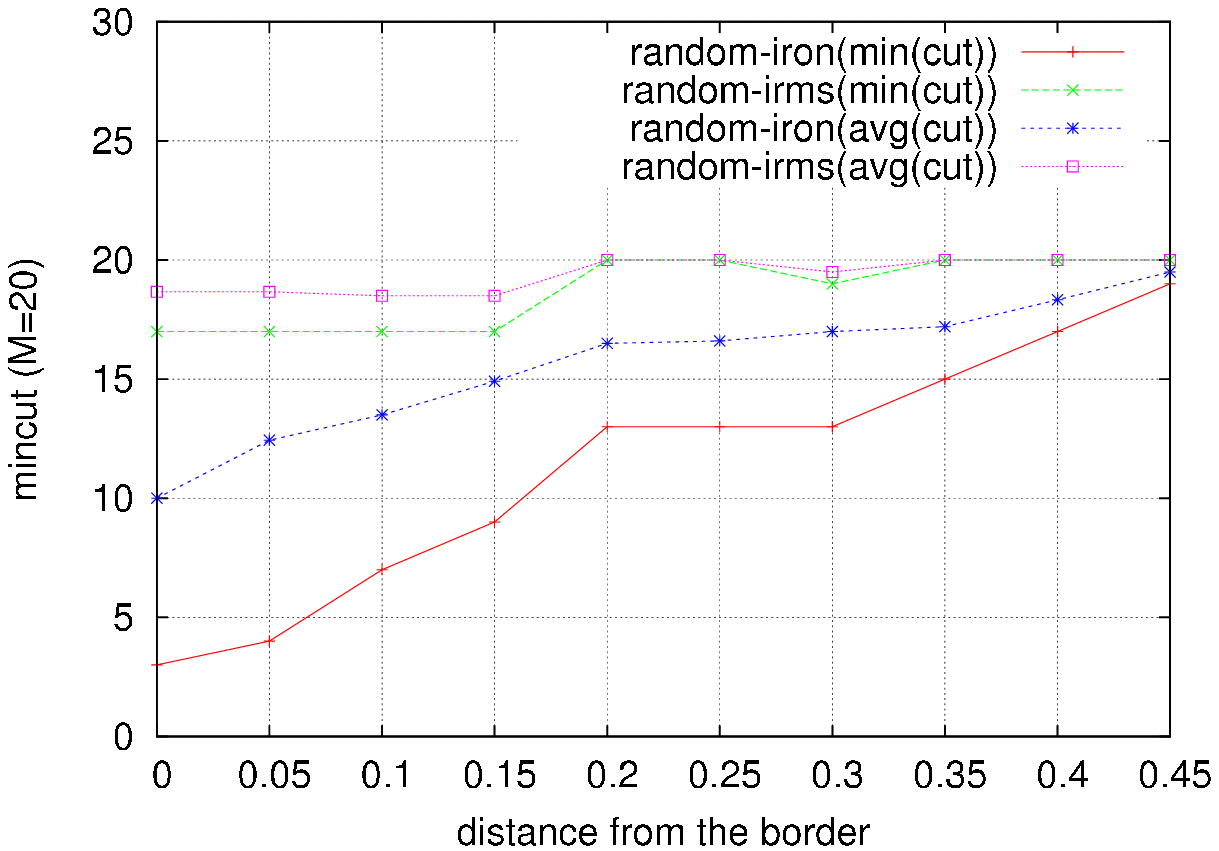}
}\hfill
%
%
\subfigure[cost vs avg neighbor num on square N=200]{\label{fig:cost-N200}
\includegraphics[width=.4\textwidth]{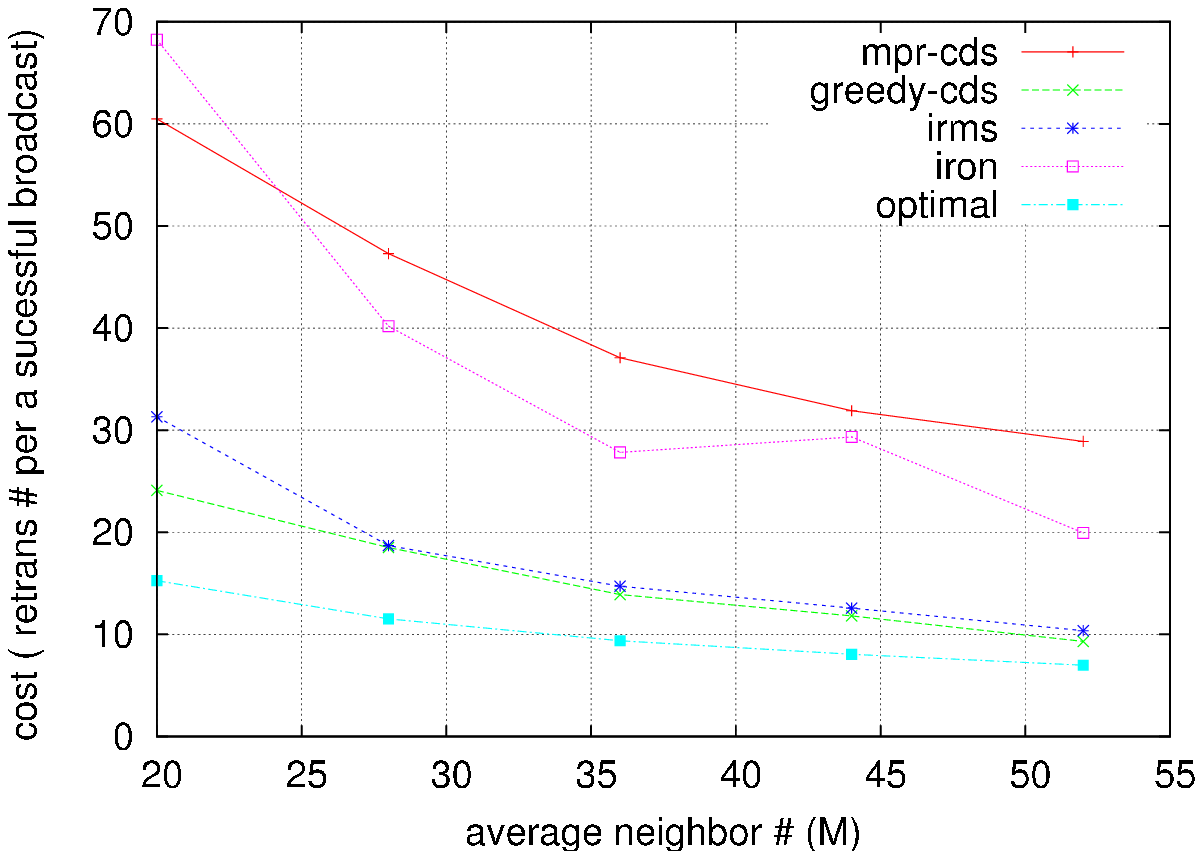}
}\hspace{.1in}
\subfigure[cost vs total node num on square M=20]{ \label{fig:cost-M20}
\includegraphics[width=.4\textwidth]{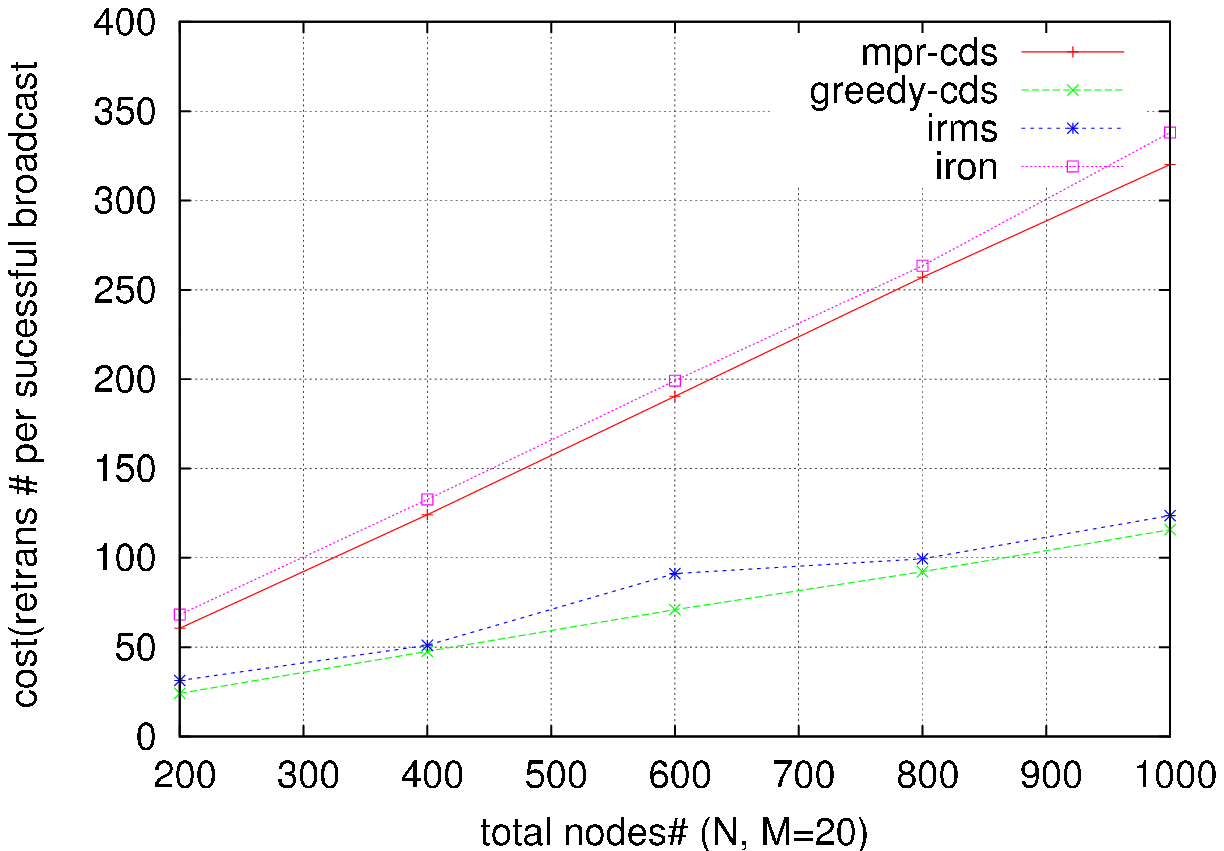}
}\hspace{.1in}
\end{figure*}
\begin{figure}
{\footnotesize\color{white}\caption{}}
\vspace{-10mm}
\end{figure}


For comparison purposes, we evaluated the IR-MS heuristic on instances of
lattice unit disk graphs and random
unit disk graphs, both with and without torus effect --- four variants
in total.
Our parameters are the following: number of nodes $N=196$,
the avg. number of neighbors is successively $4,12,28,48,80$.
\begin{figure}[!htb]
\centering
\resizebox{9cm}{!}{\includegraphics*[angle=270]{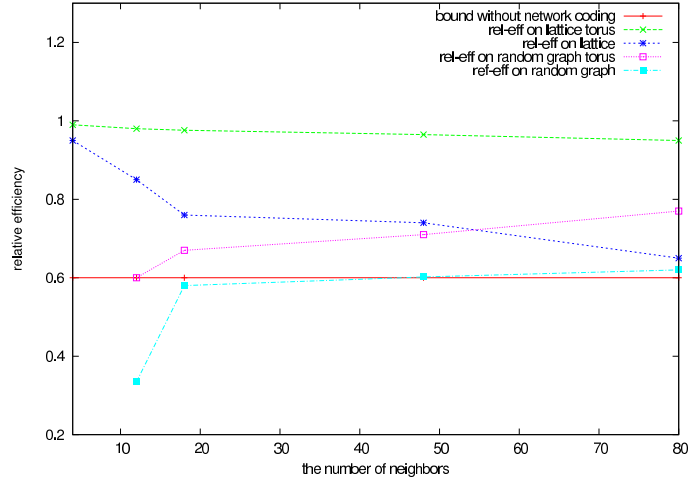}}
\caption{Relative cost (efficiency compared to optimal)}
\label{fig:efficiency}
\vspace{-3mm}
\end{figure}

\reffig{fig:efficiency} represents the
efficiency obtained for different cases
(average of 10 results). The central result is the appreciable
performance of IR-MS compared to the achievable performance
without network coding: it mostly outperforms the upper bound
without network coding ($0.609$).

Now consider the four different types of networks:\\
$\bullet$ First type, the most regular graphs: lattice unit disk
graphs on torus. IR-MS closely approaches optimality ($>0.95$). This is
because every node has the same number of neighbors and the rate
adjustment
is not needed.\\
$\bullet$ Second type, lattice unit disk graphs without torus: the
nodes near the border have less neighbors than others are
the issue. IR-MS successfully solves the border issue, 
show in next section, approaching the targeted maximum broadcast rate,
$\frac{\mathrm{min-cut}}{M} = 1$. However efficiency decreases as
density increases,
because of an increase of the cost of nodes near the border.\\
$\bullet$ Third type, random unit disk graph on torus. No border
effect here, but IR-MS has to overcome the effects of non-homogeneity,
which is done convincingly. \\
$\bullet$ Fourth type, genuine random unit disk graph. The
performance is acceptable, but on low density, performance becomes
lower, because IR-MS does not fully achieve the maximum broadcast
rate, $M$.

\subsection{Distribution of the Min-cut}

In this section, further results for case of the random unit disk graphs
are provided: these graphs were the ones with the most problematic performance.

Again, the minimum of the min-cuts $\Cmin(s,t)$ (for all $t \in \cu{V}$)
decides the overall maximum broadcast rate of
the source. Hence good performance is achieved when the
distribution of these min-cuts is tighter.
Deeper insight is gained by analyzing the cumulative
distribution of the min-cut of each node, for the random graph
$N=400$ previously studied: it is
displayed on \refsubfig{fig:mincut-cmu-dis}.
As evidenced, without IR-MS, the
distribution of the min-cut is wider, but with IR-MS, the distribution
is closer to $M$ (with a peak for $M=20$, the targetted min-cut). 
Still, there is some room for improvement, because a few nodes have
a min-cut around $15$.
%
%

The \refsubfig{fig:mincut-mu-dis} provides additional information
about the value of the min-cut depending on the position of
the nodes, for IR-MS
and IRON respectively.
The distance of each nodes to the border of the network is computed,
and statistics are made for the min-cut of nodes with 
same distance to the border. 
The \refsubfig{fig:mincut-cmu-dis} gives the average min-cut 
and the minimum min-cut for nodes at a given distance from the border.

It evidences that the border effect is key, with a lower min-cut
when the node is near the border: again, we see that
IR-MS improves the minimum min-cut but does not
always achieve the target min-cut$=M$. 

%

\subsection{Random Unit Disk Graphs $N$, $M$}

The \refsubfig{fig:cost-N200} and \refsubfig{fig:cost-M20}, show different
perpectives on the performance. The performance measured on these
graphs, is $\Ecost$, the number of retransmissions per broadcast packet.

First, different algorithms are compared:\\
$\bullet$ Network coding with IRON.\\
$\bullet$ Network coding with IR-MS.\\
$\bullet$ MPR-based dominating sets from \cite{Bib:AJV05} (with performance 
  close to MultiPoint-Relays (MPR)-based flooding of
  \cite{Bib:olsr}). It is representative of the performance
of algorithms, with only local topology information, without network coding.\\
$\bullet$ Connected dominating set from \cite{Bib:GK96} (efficient variant
  of a greedy algorithm). It is representative of the performance of
  centralized connected dominating sets algorithms 
  (and to some extent, representative
  of what could be achieved without network coding).\\
$\bullet$ for \refsubfig{fig:cost-N200}, the optimal solution 
   with network coding.

\refsubfig{fig:cost-N200} displays the performance of each algorithm
when the density of the network increases.

One result is that the performance of the optimal network coding 
is about $\frac{1}{3}$ better than the performance of the 
connected dominating set from \cite{Bib:GK96}, without network coding.

Then it appears that IR-MS is also close to this efficient dominating set,
which is already an interesting result. The gap between IR-MS and the
optimal with network coding however indicates that some improvements
are possible.

Last, MPR-based dominating sets, which use only local topology information 
are have the lowest performance: this indicates that network coding,
with IR-MS, would be useful in practice, if a fully distributed 
solution could be designed 
(with distributed min-cut computation \cite{Bib:LILI05}).

The~\refsubfig{fig:cost-M20} shows the results for increasing density ; they
are essentially similar.

In appendix~\ref{app:results}, additional results are given on 
\reffig{fig:additional}, for higher densities ($N=400$ on
\reffig{fig:cost-N400}, and $N=800$ on \reffig{fig:cost-N800}),
and for torus (torus with $N=200$ on \reffig{fig:cost-torus-N200} and 
torus with $N=400$ on \reffig{fig:cost-torus-N400}). As one can see,
the results are even better on a torus, because there is no border effect.

\subsection{Difficulties for Distributed Rate Selection}

\begin{figure}[!htb]
\vspace{-2mm}
\centering%
\resizebox{9cm}{!}{\includegraphics*{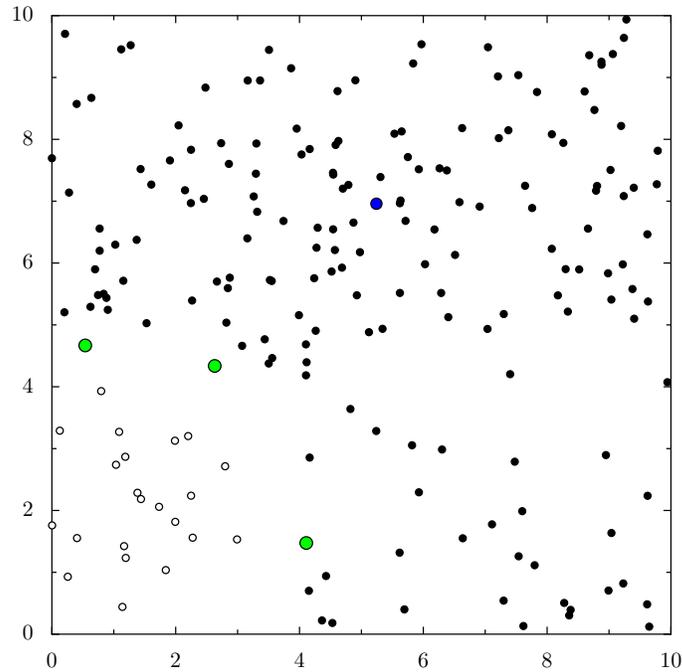}}%
\vspace{-4mm}
\caption{Example of cut}%
\label{fig:rel-eff}%
\vspace{-3mm}
\end{figure}%

Finally, the \reffig{fig:rel-eff} 
gives an example of an instance of a random unit disk graph,
where the performance of IR-MS was found to be low.
The cut ($S$/$T$ corresponding to the min-cut) for 
a node at the bottom of left corner is represented: 
the green dots are the only nodes which
are connecting the part in the corner to the rest of the network. 
With IR-MS,
the green nodes do not have extremely starving nodes as neighbors, and hence
increase their rate little. But data from the source is only transmitted 
through these green nodes to the sets of white dots, hence for this reason
their rate should be greater.
This example perfectly illustrates the
difficulties found in sparse networks: notice how coordination
between the green nodes would require multi-hop communication
to detect the issue.

\section{Conclusion}
\label{sec:conclusion}

We proposed and experimentally studied a heuristic 
for efficient broadcasting with network
coding only using static local information: one hop or two hop
neighbors. We showed excellent performance of this rate selection,
and detailed reasons for variations of performance.
Future work includes the use of dynamic information for the heuristics in
complement of the static local topology information.

\appendix

\begin{center}
  {\bf APPENDIX}
\end{center}

\section{Additional Results}
\label{app:results}

\begin{figure}[htpb]
\label{fig:additional}
\centering
\subfigure[cost vs avg neighbor num on square N=400]{\label{fig:cost-N400}
\includegraphics[width=.4\textwidth]{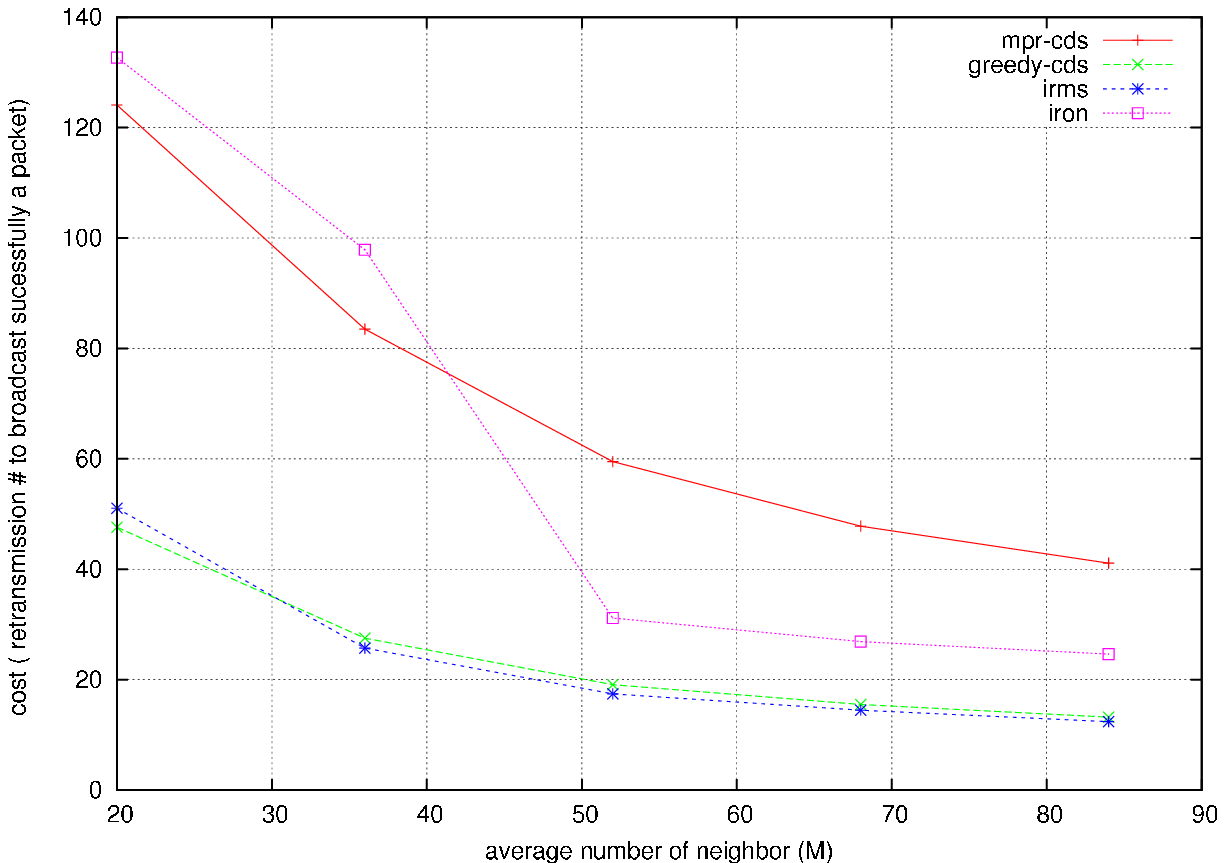}
}\hspace{.1in}
\subfigure[cost vs total node num on square N=800]{ \label{fig:cost-N800}
\includegraphics[width=.4\textwidth]{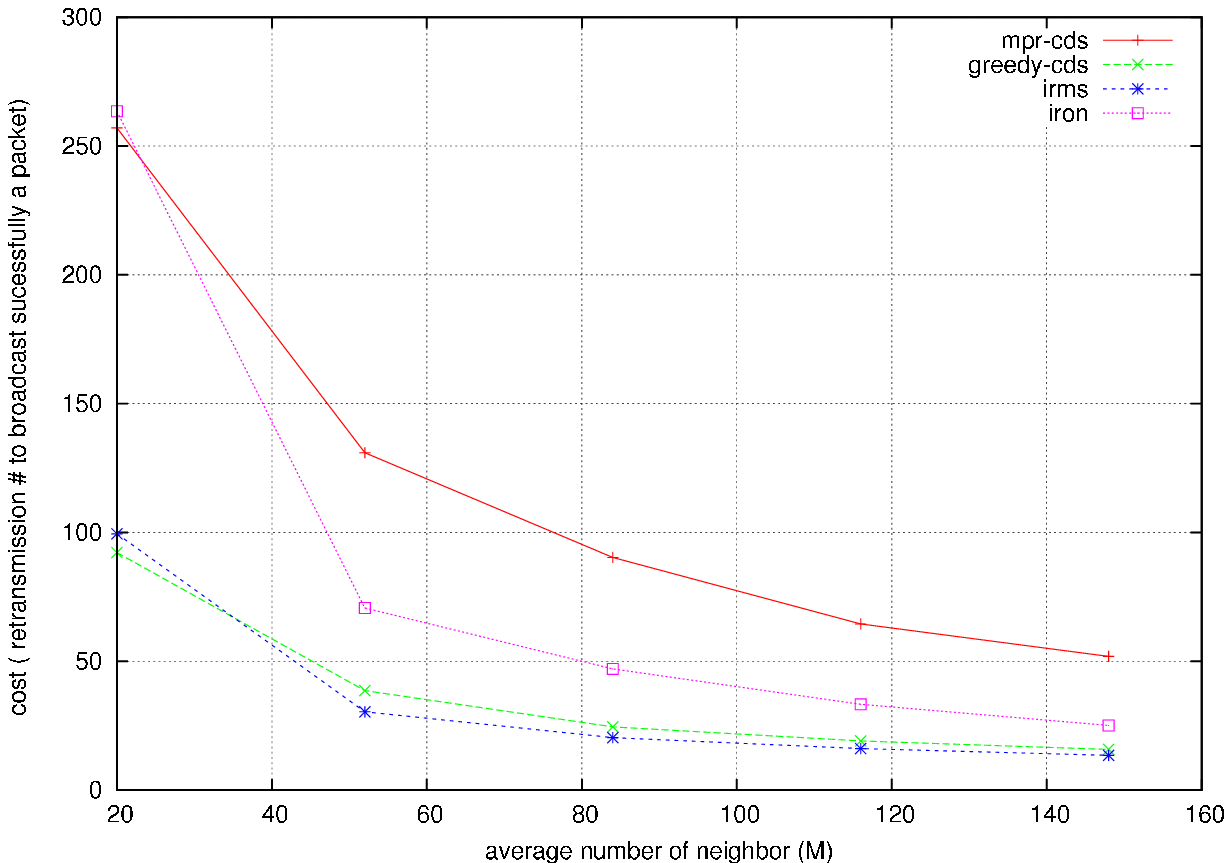}
}
\hfill
\subfigure[cost vs avg neighbor num on square {\bf torus} N=200]{\label{fig:cost-torus-N200}
\includegraphics[width=.4\textwidth]{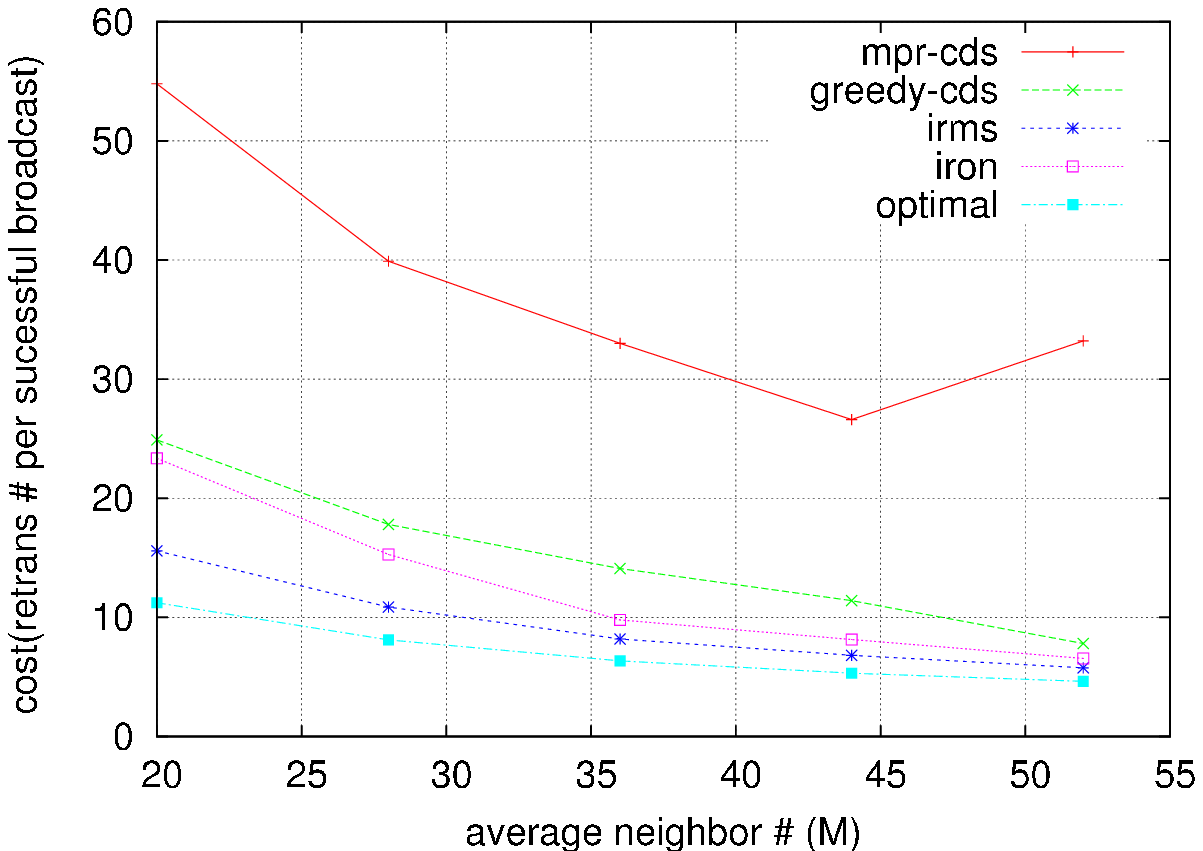}
}\hspace{.1in}
\subfigure[cost vs total node num on square M=20]{ \label{fig:cost-torus-N400}
\includegraphics[width=.4\textwidth]{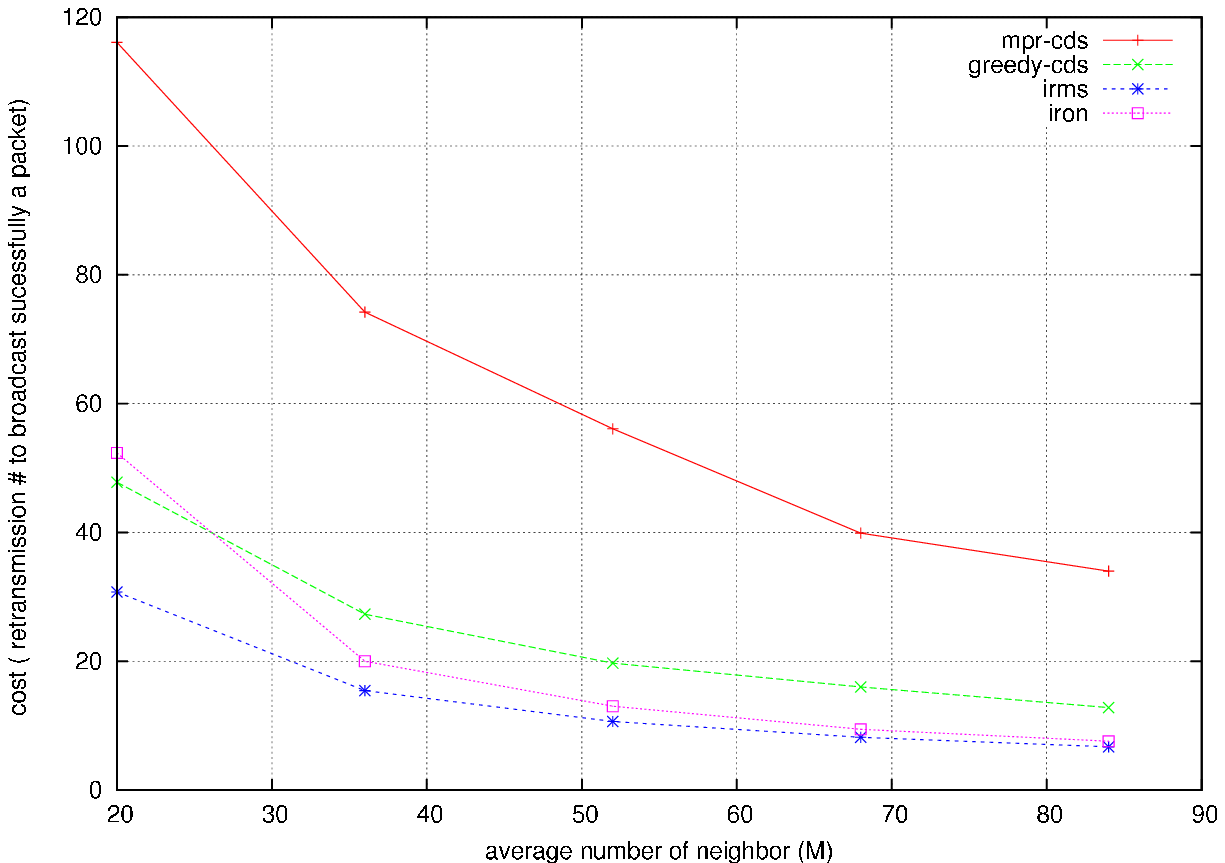}
}\hspace{.1in}
\caption{Additional results for higher density and for torus square}
\end{figure}

\end{document}